\documentclass[referee]{raa}
\usepackage{graphicx,times}
\usepackage{natbib}
\usepackage{amssymb,amsmath}
\bibpunct{(}{)}{;}{a}{}{,}

\usepackage[a4paper=true,dvipdfm=true,pagebackref=true]{hyperref}
\hypersetup{pdftitle = The title of my PDF, pdfauthor = My name,
pdfsubject= The subject, pdfkeywords = keyword1 keyword2 keyword3}
\hypersetup{colorlinks = true, linkcolor = green, anchorcolor = red,
citecolor = blue, filecolor = red, pagecolor = red, urlcolor = red}

\usepackage{subfigure}

\begin{document}

\title{Transient acceleration in $f(T)$ gravity$^*$ \footnotetext{\small $*$ Supported by the National Natural
Science Foundation of China.}}

   \volnopage{Vol.0 (200x) No.0, 000--000}      
   \setcounter{page}{1}          

   \author{Jing-Zhao Qi
      \inst{1}
   \and Rong-Jia Yang
      \inst{2}
   \and Ming-Jian Zhang
      \inst{1}
   \and Wen-Biao Liu
      \inst{1}
   }

   \institute{Department of Physics, Institute of Theoretical
Physics, Beijing Normal University, Beijing, 100875, China; {\it wbliu@bnu.edu.cn}\\
        \and
             College of Physical Science and Technology,
Hebei University, Hebei Baoding, 071002, China\\
\vs \no
   {\small Received 2015 X. X.; accepted 2015 X. X.}
}

\abstract{Recently a $f(T)$ gravity based on the modification of the teleparallel gravity was proposed to explain the accelerated expansion of the universe without the need of dark energy.
We use observational data from Type Ia Supernovae, Baryon Acoustic Oscillations, and Cosmic Microwave Background to constrain this $f(T)$ theory and reconstruct the effective equation of state and the deceleration parameter. We obtain the best-fit values of parameters and find an interesting result that the constrained $f(T)$ theory allows for the accelerated Hubble expansion to be a transient effect.
 }

\authorrunning{J.-Z. Qi et al. }            
\titlerunning{Transient acceleration in $f(T)$ gravity}  

\maketitle

\section{Introduction}
\label{introduction}

 A series of independent cosmological observations including the type Ia supernovae (SNIa) \citep{riess1998supernova}, large scale structure \citep{tegmark2004cosmological}, baryon acoustic oscillation (BAO) peaks \citep{eisenstein2005detection} and cosmic microwave background (CMB) anisotropy \citep{spergel2003wmap} have probed the accelerating expansion of the universe. Subsequently, many gravitational theories and cosmological models have been proposed to explain this cosmological phenomenon. Under the assumption of cosmological principles, these theories include the mysterious dark energy with negative pressure in general relativity and modify gravity models to the general relativity. For the former, the acceleration is realized by the drive of exotic dark energy, such as the cosmological constant, quintessence or phantom. The cosmological constant model ($\Lambda$CDM) is the simplest candidate for dark energy models, and agrees well with current cosmological observations. However, the $\Lambda$CDM model is faced with the fine-tuning problem \citep{weinberg1989cosmological} and coincidence problem \citep{zlatev1999quintessence}. Moreover, the nature of dark energy in form of other candidates still cannot be revealed. For the latter, the acceleration is realized by modification to the general relativity without exotic dark energy, such as the brane-world Dvali-Gabadadze-Porrati model \citep{dvali2000metastable}, $f(R)$ gravity \citep{chiba20031}, Gauss-Bonnet gravity \citep{nojiri2005modified}.

Similar as the exotic dark energy and other modified gravity models, it is found that the cosmic acceleration can also be obtained successfully from another gravitational scenario described by the $f(T)$ theory \citep{bengochea2009dark}. Proposed based on the Teleparallel Equivalent of General Relativity (also known as Teleparallel Gravity), scalar $T$ is the Lagrangian of teleparallel gravity. The teleparallel gravity is not a new theory of gravity, but an alternative geometric formulation of the general relativity. In teleparallel gravity, the Levi-Civita connection used in Einstein's general relativity is replaced by the Weitzenb\"ock connection with torsion. However, the torsion vanishes in the dark energy and modified gravity models. Moreover, $f(T)$ theories have several interesting features: they not only can explain the late accelerating expansion, but also have second order differential equations, which are simpler than the $f(R)$ gravity. In addition, when certain conditions are satisfied, the behavior of $f(T)$ will be similar to quintessence \citep{xu2012phase}. Although $f(T)$ gravity has attracted wide attention, a disadvantage pointed out in Ref. \citep{li2011f} is that the action and the field equations of $f(T)$ do not respect local Lorentz symmetry. Nonetheless, the $f(T)$ gravity might provide a significant alternative to conventional dark energy in general relativistic cosmology. In addition, the Ref. \citep{2011JCAP...03..046S} indicated that the Lorentz invariance violation is still possible, while $f(T)$ gravity might provide some insights about Lorentz violation. Such $f(T)$ theories are worth further depth studies.

Up to now, a number of $f(T)$ theories have been proposed \citep{bengochea2009dark,linder2010einstein,
yang2011new,myrzakulov2011accelerating,bamba2011equation,wu2011f}. Under these cases, Yang found  that $f(T)$ theories are not dynamically equivalent to teleparallel action with an added scalar field \citep{yang2011conformal}. Like other gravity theories and models, the $f(T)$ theories also have been investigated using the popular observational data. Investigations show that the $f(T)$ theories are compatible with observations (see e.g. \citep{nesseris2013viable,Zheng2010am} and references therein). We note that the new type of $f(T)$ theory was proposed to explain the accelerating expansion of the universe, and it behaves like a cosmological constant; but because of its dynamic behavior, it is free from the coincidence problem seen in the case of $\Lambda$CDM \citep{yang2011new}. Due to this characteristic, this type of $f(T)$ is possible to be distinguished from a $\Lambda$CDM model. However, observational analysis for this model is still absent. Hence, we would like to perform some further analysis using the observational data, such as the SNIa, BAO, and CMB.

This paper is organized as follows. In Sec \ref{fT theory}, the general $f(T)$ gravity and the $f(T)$ model proposed in \citep{yang2011new} are introduced. In Sec \ref{data and method}, we describe the method for constraining cosmological models and reconstruction scheme. Subsequently, the parameters of the specific $f(T)$ model are constrained by observational data. Further more, through the reconstruction scheme the effective equation of state and the deceleration parameter are reconstructed in Sec \ref{result}. Finally, we give the summary and conclusions in Sec \ref{conclusion}.

\section{The $f(T)$ theory}
\label{fT theory}

The $f(T)$ theory is a modification of teleparallel gravity, which uses the curvatureless Weitzenb\"ock connection instead of torsionless Levi-Civita connection in Einstein's General Relativity. The curvatureless torsion tensor is
\begin{equation}
\label{torsion2}
T^\lambda_{\:\:\mu\nu}\equiv e^\lambda_i(\partial_\mu e^i_\nu-\partial_\nu e^i_\mu),
\end{equation}
where $e^\mu_i$ $(\mu=0,1,2,3)$ are components of the four linearly independent vierbein field ${\mathbf{e}_i(x^\mu)}$ ($i=0, 1, 2, 3$) in a coordinate basis. In particular, the vierbein is an orthonormal basis for the tangent space at each point $x^\mu$ of the manifold: $\mathbf{e}
_i\cdot\mathbf{e}_j=\eta_{i\, j}$, where $\eta_{i\, j}=$diag $(1,-1,-1,-1)$. Notice that Latin indices refer to the tangent space, while Greek indices label coordinates on the manifold. The metric tensor is obtained from the
dual vierbein as $g_{\mu\nu}(x)=\eta_{i\, j}\, e^i_\mu (x)\, e^j_\nu (x)$.
The torsion scalar is the Lagrangian of teleparallel gravity  \citep{bengochea2009dark}
\begin{equation}  \label{lagTele}
T\equiv S_\rho^{\:\:\:\mu\nu}\:T^\rho_{\:\:\:\mu\nu},
\end{equation}
where
\begin{equation}  \label{S}
S_\rho^{\:\:\:\mu\nu}=\frac{1}{2}\Big(K^{\mu\nu}_{\:\:\:\:\rho}+\delta^\mu_%
\rho \:T^{\theta\nu}_{\:\:\:\:\theta}-\delta^\nu_\rho\:
T^{\theta\mu}_{\:\:\:\:\theta}\Big),
\end{equation}
and the contorsion tensor $K^{\mu\nu}_{\:\:\:\:\rho}$ is given by
\begin{equation}  \label{K}
K^{\mu\nu}_{\:\:\:\:\rho}=-\frac{1}{2}\Big(T^{\mu\nu}_{\:\:\:\:\rho}
-T^{\nu\mu}_{\:\:\:\:\rho}-T_{\rho}^{\:\:\:\:\mu\nu}\Big).
\end{equation}
In the $f(T)$ theory, we allow the Lagrangian density to be a function of $T$ \citep{bengochea2009dark,ferraro2007modified,linder2010einstein}, thus the action reads
\begin{equation}  \label{accionTP}
I= \frac{1}{16\, \pi\, G}\, \int d^4x\:e\:f(T),
\end{equation}
where $e=det(e^i_\mu)=\sqrt{-g}$. The corresponding field equation is
\begin{eqnarray}
[e^{-1}\partial_\mu(e\:S_i^{\:\:\:\mu\nu})-e_i^{\:\lambda}
\:T^\rho_{\:\:\:\mu\lambda}\:S_\rho^{\:\:\:\nu\mu}]f_T+ S_i^{\:\:\:\mu\nu}\partial_\mu T f_{TT}
+\frac{1}{4}
\:e_i^\nu \:f(T)=\frac{1}{2}k^2\:e_i^{\:\:\:\rho}\:T_\rho^{\:\:\:\nu},  \label{ecsmovim}
\end{eqnarray}
where $k^2=8\pi G$, $f_T\equiv df/dT$, $f_{TT} \equiv d^2f/dT^2$,  $S_i^{\:\:\mu\nu}\equiv e_i^{\:\:\rho}S_\rho^{\:\:\mu\nu}$, and
$T_{\mu\nu}$
is the matter energy-momentum tensor. Obviously, Eq.(\ref{ecsmovim}) is a second-order equation. Thus, the $f(T)$ theories are simpler than the $f(R)$ theories with fourth-order equations.

Considering a flat homogeneous and isotropic FRW universe, we have
\begin{equation}  \label{tetradasFRW}
e^i_\mu={\rm diag}\left(1,a(t),a(t),a(t)\right)~,~~~~
e^\mu_i={\rm diag}\left(1,\frac{1}{a(t)},\frac{1}{a(t)},\frac{1}{a(t)}\right),
\end{equation}
where $a(t)$ is the cosmological scale factor. By substituting Eqs.(\ref{tetradasFRW}), (\ref{torsion2}), (\ref{S}) and (\ref{K}) into Eq. (\ref{lagTele}), we obtain the torsion scalar \citep{bengochea2009dark}
\begin{equation}  \label{STFRW}
T\equiv S^{\rho\mu\nu}T_{\rho\mu\nu}=-6\:H^2,
\end{equation}
where $H$ is the Hubble parameter $H=\dot{a}/a$. The dot represents the first derivative with respect to the cosmic time. Substituting Eq. (\ref{tetradasFRW}) into (\ref{ecsmovim}), one can obtain the corresponding Friedmann equations
\begin{eqnarray}\label{f1}
12H^2 f_{T}+f=2k^2 \rho ,\\
\label{f2}
48 H^2 \dot{H}f_{TT}-(12H^2+4\dot{H})f_{T}-f=2k^2p,
\end{eqnarray}
with $\rho$ and $p$ as the total energy density and pressure, respectively. The detailed calculation can be found in Ref. \citep{bengochea2009dark}. The conservation equation reads
\begin{eqnarray}
\label{rho}
\dot{\rho}+3H(\rho+p) &=& 0.
\end{eqnarray}
We should note that the only components considered here are matter and radiation, but not dark energy. After brief simplification to the Friedmann Eqs.(\ref{f1}) and (\ref{f2}), we can rewrite them as
\begin{eqnarray}
\frac{3}{k^2}H^2 &=& \rho+\rho_{\rm eff}, \\
\frac{1}{k^2}(2\dot{H}+3H^2) &=& -(p + p_{\rm eff}),
\end{eqnarray}
where the effective energy density $\rho_{\rm eff}$ and pressure $p_{\rm eff}$ contributed from torsion are respectively given by \citep{yang2011new}
\begin{eqnarray}
\label{rhoT}
\rho_{\rm eff} &=& \frac{1}{2k^2}(-12H^2f_T-f+6H^2), \\
\label{pT}
p_{\rm eff} &=& -\frac{1}{2k^2}[48\dot{H}H^2f_{TT}-4\dot{H}f_{T}+4\dot{H}] - \rho_{\rm eff}.
\end{eqnarray}
We term it ``effective" because it is just a geometric effect instead of a specific cosmic component. Therefore, what we are interested in is the acceleration driven by the torsion, not the exotic dark energy. Using Eqs.(\ref{rhoT}) and (\ref{pT}), we can define the total and effective equation of state as  \citep{yang2011new}
\begin{eqnarray}
w_{\rm tot} & \equiv & \frac{p + p_{\rm eff}}{\rho + \rho_{\rm eff}}=-1+\frac{2(1+z)}{3H}\frac{dH}{dz},\\
w_{\rm eff} & \equiv & \frac{p_{\rm eff}}{\rho_{\rm eff}}=-1-\frac{48\dot{H}H^2f_{TT}-4\dot{H}f_{T}+4\dot{H}}{-12H^2f_T-f+6H^2}.
\end{eqnarray}
The deceleration parameter, as usual, is defined as
\begin{eqnarray}
q(z) \equiv -\frac{\ddot{a}}{aH^2}=-1+\frac{(1+z)}{H}\frac{dH}{dz}.
\end{eqnarray}

After reviewing the general formation of $f(T)$ gravity, we now focus on a type of $f(T)$ gravity proposed in Ref. \citep{yang2011new}
\begin{equation}
\label{ft}
f(T)=T-\alpha T_0 \big[\big(1+\frac{T^2}{T^2_0} \big)^{-n}-1 \big],
\end{equation}
which is analogue with a type of $f(R)$ theory proposed in Ref. \citep{starobinsky2007disappearing}, where $\alpha$ and $n$ are positive constants. $T_0=-6H^2_0$ and $H_0$ is the current value of the Hubble parameter. This type of $f(T)$ gravity has attracted much attention and been discussed in detail in Ref. \citep{Sharif2013kka}. Here we will look into the observational constraints on this type of $f(T)$ gravity. With $f(T)$ taking the form of Eq. (\ref{ft}), Eq.(\ref{f1}) can be rewritten as
\begin{eqnarray}
\label{y2}
E^2+\frac{4n\alpha E^4}{(1+E^4)^{n+1}}+\frac{\alpha}{(1+E^4)^{n}}  -B=\alpha,
\end{eqnarray}
where $E^2\equiv H^2/H^2_0$ and $B=\Omega_{m0}(1+z)^3$, with $\Omega_{m0}$ being the matter density parameter today. Here we only focus on the evolution of the universe at low redshift, so we neglect the contribution of radiation. For $E(z=0)=1$, we have $\alpha=(1-\Omega_{m0})/(1-2^{-n+1} n-2^{-n})$. This $f(T)$ model has some interesting characteristics: firstly, the cosmological constant is zero in the flat space-time because $f(T=0)=0$, while the geometrical one attributes as the dark energy; secondly, it can behave like the cosmological constant. Such characteristics indicate that this type of $f(T)$ model is possible to be accepted by observational data, while impossible to be distinguished from the $\Lambda$CDM. Moreover, though the behavior of this type of $f(T)$ theory is similar to $\Lambda$CDM because of its dynamic behavior, it can avoid the coincidence problem suffered by $\Lambda$CDM.

\section{Observational data and fitting method}
\label{data and method}

In this section, we would like to introduce the observational data and constraint method. The corresponding observational data here are distance moduli of SNIa, CMB shift parameter and BAO distance parameter.

\subsection{Type Ia supernovae}  \label{SNIa data}

As early as 1998, cosmic
accelerating expansion was first observed by "standard candle"
SNIa which has the same intrinsic luminosity. Therefore, the
observable is usually presented in the distance modulus, the
difference between the apparent magnitude $m$ and the absolute
magnitude $M$. The latest version is Union2.1 compilation which includes 580 samples \citep{suzuki2012hubble}. They are
discovered by the Hubble Space Telescope Cluster Supernova Survey
over the redshift interval $0.01<z<1.42$. The theoretical
distance modulus is given by
\begin{equation}
\mu_{th}(z)= m-M= 5 \log_{10}D_L(z)+\mu_0,
\end{equation}
where $\mu_0=42.38-5\log_{10}h$, and $h$ is the Hubble
constant $H_0$ in the units of 100 km s$^{-1}$Mpc$^{-1}$. The
corresponding luminosity distance function $D_L(z)$  is
\begin{eqnarray}\label{dl}
 D_L(z) = (1 + z) \int_{0}^{z} \frac{dz'}{E(z'; \textbf{p})},
\end{eqnarray}
where $E(z'; \textbf{p})$ is the dimensionless Hubble parameter given by Eq.(\ref{y2}), and \textbf{p} stands for the parameter vector of the evaluated model embedded in the expansion rate parameter $E(z)$. We note that parameters in the expansion rate $E(z)$ include the annoying parameter $h$. In order to exclude the Hubble constant, we should marginalize over the nuisance parameter $\mu_0$  by integrating the probabilities on $\mu_0$ \citep{pietro2003future,nesseris2005comparison,perivolaropoulos2005constraints}. Finally, we can estimate
the remaining parameters by minimizing
\begin{equation} \label{chi2_SN2}
    \tilde{\chi}^{2}_{\rm SN}(z,\textbf{p})= A - \frac{B^2}{C},
\end{equation}
where
\begin{eqnarray}
     A(\textbf{p}) &=&  \sum_{i} \frac{[\mu_{obs}(z) - \mu_{th}(z; \mu_0 = 0,
     \textbf{p})]^2}{\sigma_{i}^{2}(z)},   \nonumber\\
     B(\textbf{p}) &=&  \sum_{i} \frac{\mu_{obs}(z) - \mu_{th}(z; \mu_0=0,
     \textbf{p})}{\sigma_{i}^{2}(z)},   \nonumber\\
     C &=&  \sum_{i} \frac{1}{\sigma_{i}^{2}(z)} , \nonumber
\end{eqnarray}
and $\mu_{obs}$ is the observational distance modulus. This approach has been used in the reconstruction of dark energy
\citep{wei2007reconstruction}, parameter constraint
\citep{wei2010observational},  reconstruction of the energy
condition history \citep{wu2012reconstructing} etc.

\subsection{Cosmic microwave background}  \label{cmb data}

The CMB experiment measures the temperature and
polarization anisotropy of the cosmic radiation in early
epoch. It generally plays a major role in establishing and
sharpening the cosmological models. In the CMB measurement, the shift parameter
$R$ is a convenient way to quickly evaluate the likelihood
of the cosmological models, and contains the main information of the CMB observation \citep{hu1996small,hinshaw2009five}. It is expressed as
\begin{equation}
R=\sqrt{\Omega_{m0}}\int^{z_s}_0\frac{dz'}{E(z';\textbf{p})},
\end{equation}
where $z_s=1090.97$ is the redshift of decoupling.  According to the measurement of
WMAP-9 \citep{hinshaw2012nine}, we estimate the parameters by minimizing the corresponding
$\chi^2$ statistics
    \begin{equation}  \label{cmb constraint}
     \chi ^2_R = \left( \frac{R-1.728}{0.016} \right)^2.
     \end{equation}

\subsection{Baryon acoustic oscillation}   \label{bao data}

The measurement of BAO in the large-scale galaxies has rapidly become one of the most important observational pillars in cosmological constraints. This measurement is usually called the standard ruler in cosmology \citep{eisenstein1998baryonic}.
The distance parameter $A$ obtained from the BAO peak in the distribution of SDSS luminous red galaxies \citep{eisenstein2005detection} is a significant parameter and defined as
\begin{equation}
A_{th}=\Omega_{m0}^{1/2}E(z_1)^{-1/3}\left[\frac{1}{z_1}\int^{z_1}_{0}\frac{dz'}{E(z';\textbf{p})}\right]^{2/3}.
\end{equation}
We use the three combined data points in Ref. \citep{addison2013cosmological} that cover $0.1 < z <2.4$ to determine the parameters in evaluated models. The expression of $\chi^2$ statistics is
\begin{equation}
\chi^2_{\rm{A}}=\sum_{i} \left(\frac{A_{th}-A_{obs}}{\sigma^2_A}\right)^2,
\end{equation}
where $A_{obs}$ is the observational distance parameter and $\sigma_A$ is its corresponding error.

Since the SNIa, CMB, and BAO data points are effectively
independent measurements, we can simply minimize their total
$\chi^{2}$ values
\begin{eqnarray}
\label{21}\chi^2(z, \textbf{p})= \tilde{\chi}^{2}_{\rm SN} + \chi^{2}_{\rm R}+\chi^{2}_{\rm
A}, \nonumber
\end{eqnarray}
to determine the parameters in the evaluated $f(T)$ model.

\subsection{Reconstructing method}  \label{reconstruction method}

Using the above introduced $\chi^2$ statistics, we can obtain the best-fit values and their errors of basic parameters $\textbf{p}$. Further, we can reconstruct the other variable $F$ relative with the known basic parameters $\textbf{p}$ by error propagation following the method in Ref. \citep{lazkoz2012first}. For example, the estimation from the observational data on the $i$th parameter $p_i$ is  $p_i={p_{0i}}^{+\sigma_{iu}}_{-\sigma_{il}}$, where $p_{0i}$ is the best-fit value, $\sigma_{iu}$ and $\sigma_{il}$ are the upper limit and lower limit, respectively. Errors of the reconstructed function $F$ are estimated by
    \begin{eqnarray}   \label{construct}
    \delta F_u &=& \sqrt{\sum_{i} \left[ \textrm{max}\Big( \frac{\partial F}{\partial p_i}\sigma_{iu}, -\frac{\partial F}{\partial p_i}\sigma_{il} \Big)
    \right]^2},  \nonumber\\
    \delta F_l &=& \sqrt{\sum_{i} \left[ \textrm{min}\Big( \frac{\partial F}{\partial p_i}\sigma_{iu}, -\frac{\partial F}{\partial p_i}\sigma_{il} \Big)
    \right]^2},
    \end{eqnarray}
where $\delta F_u$ and $\delta F_l$ are its upper and lower bound, respectively. In this paper, we will use this method to reconstruct the effective equation of state $w_{\rm eff}$ and deceleration parameter $q$.

\section{Constraint result}  \label{result}

\begin{figure}
\begin{center}
\includegraphics[scale=0.5]{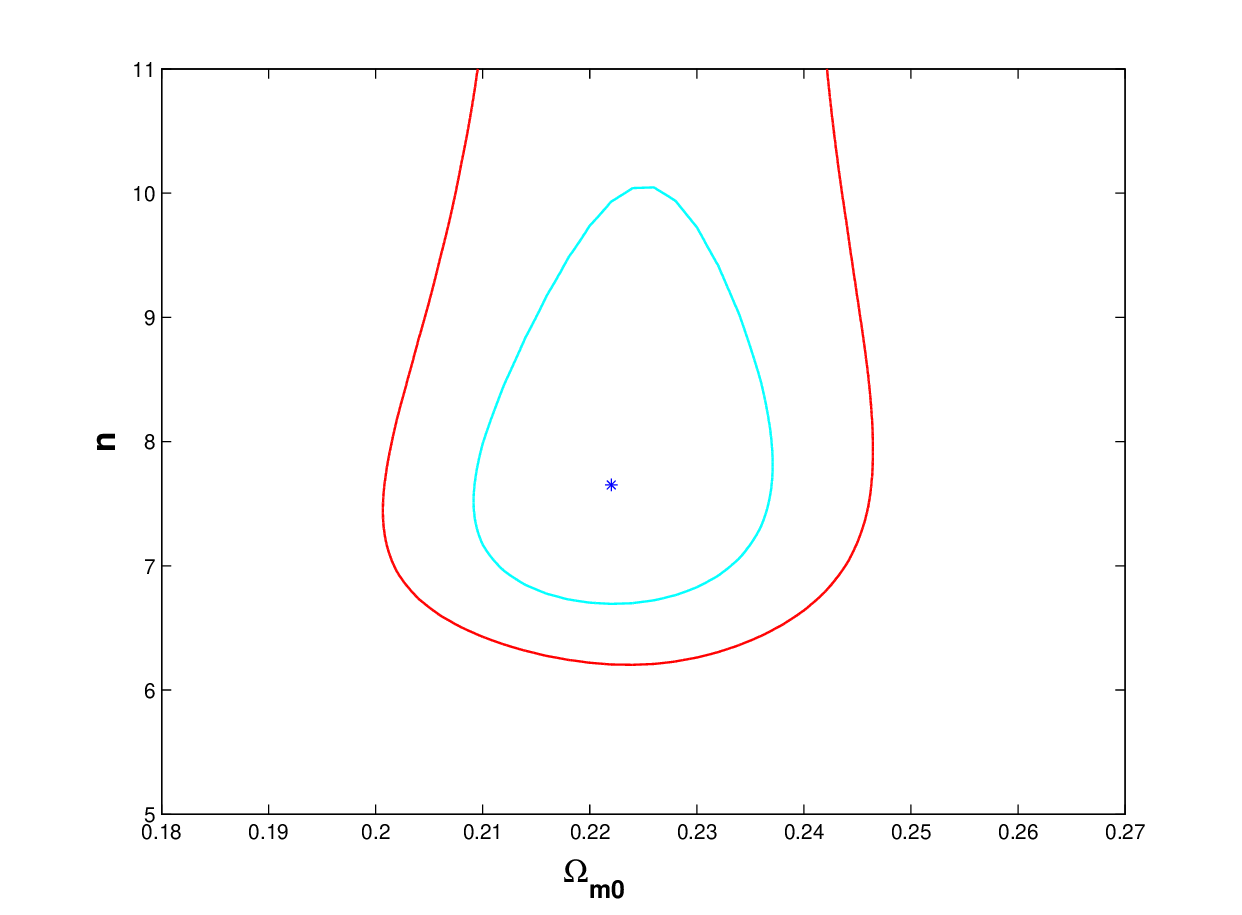}
\caption{Constraints on $f(T)$ theory with $68.3\%$ and $95.4\%$ confidence regions in the $\Omega_{m0}$-$n$ plane fitted on combinational observation of SNIa, BAO, and CMB data. The blue asterisk is the best-fit point. \label{fig1}}
\end{center}
\end{figure}

Using the observational data sets, we perform the $\chi^2$ statistics and display the contour constraint in Figure \ref{fig1}. We find that the combined data gives mild constraints on them, i.e., $\Omega_{m0}=0.22^{+0.0089}_{-0.0094}  (1\sigma)$ and $n=7.64^{+1.1750}_{-0.6700} (1\sigma)$ with $\chi^2_{\rm min}=579.4786$. If we consider the degrees of freedom (dof) $\chi^2_{\rm
min}/$dof=0.9923 indicating that this $f(T)$ model is well consistent with the observations. However, we note that the parameter $n$ is worse at $95.4\%$ confidence level. Namely, $n$ is larger than 6. If the parameter $n$ approaches infinity, we find from Eq. (\ref{ft}) that this $f(T)$ model eventually evolves to the standard $\Lambda$CDM model.

\begin{figure}
\begin{center}
\includegraphics[width=10cm]{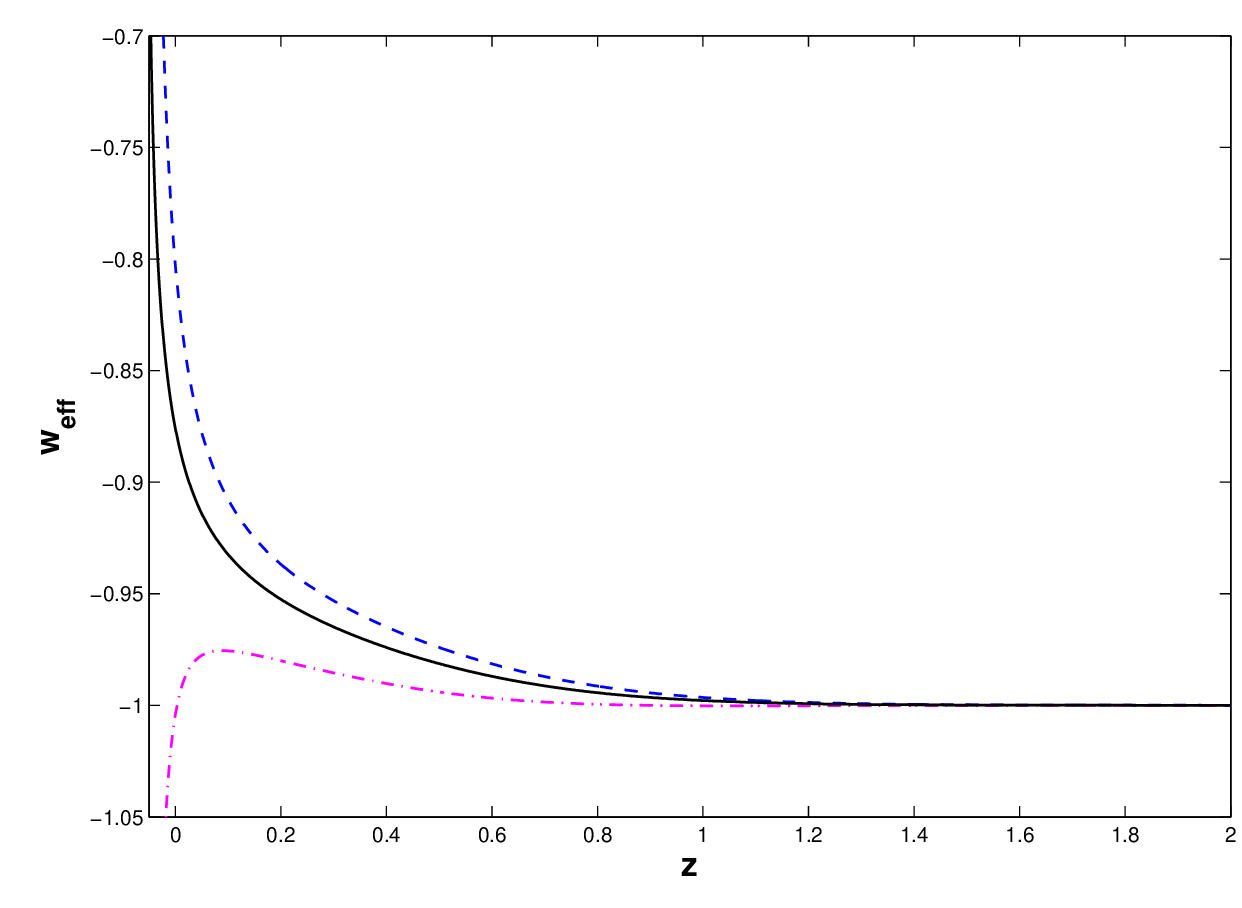}
\caption{Reconstruction of the effective equation of state at $68.3\%$ confidence level for the $f(T)$ gravity considered here. \label{fig2}}
\end{center}
\end{figure}

\begin{figure}
\begin{center}
\includegraphics[width=10cm]{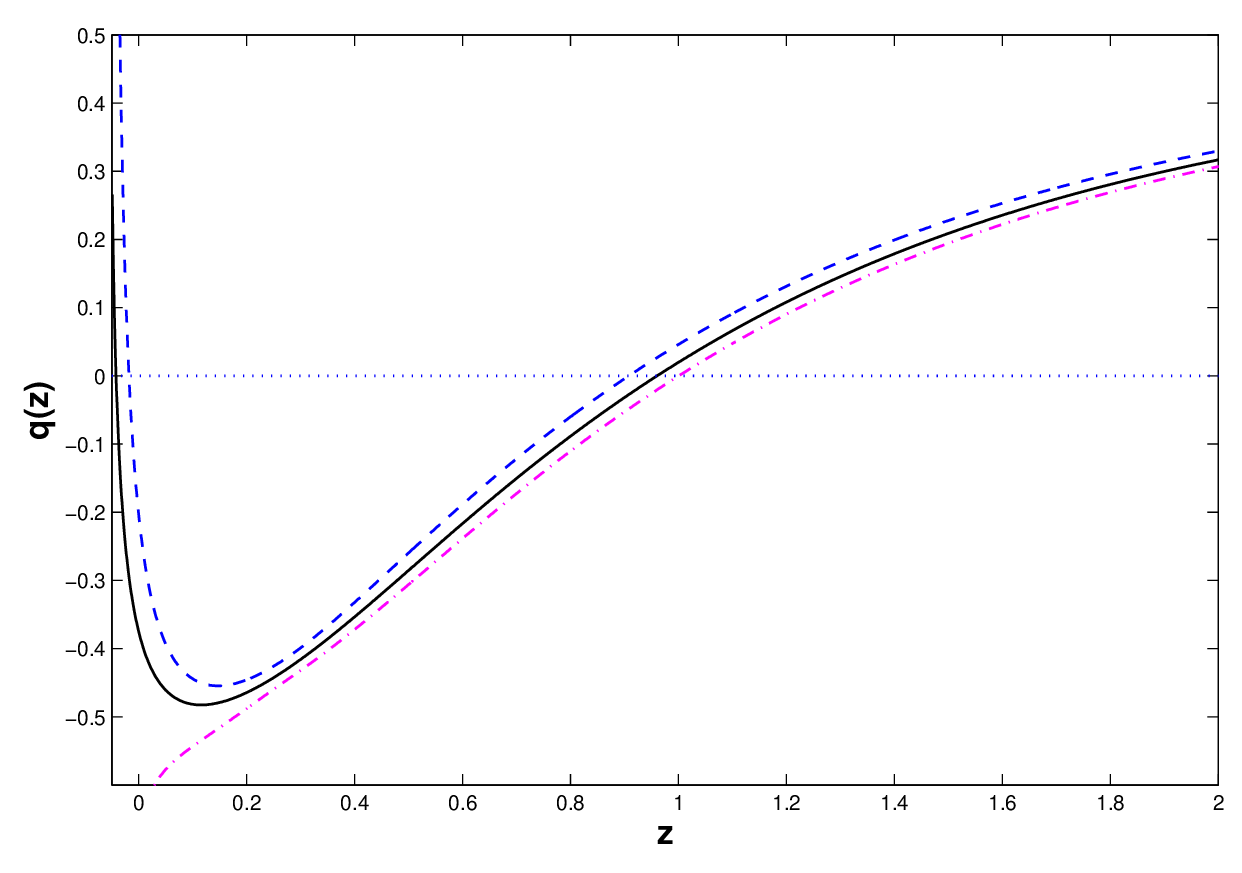}
\caption{Reconstruction of the deceleration parameter at $68.3\%$ confidence level for the $f(T)$ gravity considered here. \label{fig3}}
\end{center}
\end{figure}

In terms of Eq. (\ref{construct}), we reconstruct the effective equation of state in Figure \ref{fig2}. We find that $w_{\rm{eff}}(z)$ is a decreasing function of redshift, and steadily approaches to -1 for high redshift $z \gtrsim 1$. That is, the geometric effect behaves like the cosmological constant at early epoch. However, it generally increases with the decrease of redshift. The present value of the effective equation of state finally reaches $w_{\rm{eff0}}=-0.8760$. Moreover, the $w_{\rm{eff}}(z)$  crosses through -1 for $z<1$ within $1\sigma$ confidence level. In Figure \ref{fig3}, we also reconstruct the deceleration parameter $q(z)$. We find that the transition from decelerating to accelerating expansion occurs at $z=0.95 \pm 0.05$, which is earlier than some phenomenological deceleration parameters \citep{riess2004type,cunha2008transition}. With the decrease of deceleration parameter, its value today is $q_0=-0.3750$. In the near future $z=-0.04$, the $q(z)$ crosses the zero. That is to say, the accelerating expansion of the universe may be slowing down again and till to decelerating expansion take place in future. It is possible, however, to have an eternal accelerated phase at 68.3\% confidence level as shown in Figure \ref{fig3}. The feature of transient acceleration makes this $f(T)$ gravity compatible with the S-matrix description of string theory \citep{Banks:2001ze,Hellerman:2001yi}.
Most of dark energy models including the current standard $\Lambda$CDM scenario predict an eternally accelerating universe. But the consequent cosmological event horizon dose not allow the construction of a conventional S-matrix to describe particle interactions \citep{Guimaraes:2010mw}. However, from the standpoint of string theory, the existence of conventional S-matrix is absolutely essential for an asymptotically large space at infinity \citep{2013RAA....13..629C}. Therefore, S-matrix is ill-defined in an eternal accelerating universe. In order to alleviated the conflict between dark energy and String theory, several dynamic dark energy models have been proposed to achieve the possibility of transient acceleration phenomenon \citep{2013RAA....13..629C,2004PhLB..600..185R,carvalho2006scalar}. In addition, recently it was also argued that the SNIa data favors a transient acceleration \citep{Shafieloo:2009ti}which is not excluded by current observations \citep{Guimaraes:2010mw,Vargas:2011sz,Bassett:2002qu}. Our result indicates that this type of $f(T)$ gravity serves as an alternative from modification of gravity to the dynamic dark energy models.

\section{Summary and conclusions}   \label{conclusion}

The $f(T)$ gravity based on modification of the teleparallel gravity was proposed to explain the accelerating expansion of the universe without the need of dark energy. A brief overview of a specific $f(T)$ gravity proposed in \citep{yang2011new} was also given. We also introduced the method used to constrain cosmological models with observational data including SNIa, BAO, and CMB. After constraining the $f(T)$ gravity proposed in \citep{yang2011new}, we find that the best-fit values of the parameters at the $68.3\%$ confidence level are: $\Omega_{
m0}=0.22^{+0.0089}_{-0.0094}$ and $n=7.64^{+1.175}_{-0.67}$ with $\chi^2_{\rm min}=579.4786$ ($\chi^2_{\rm
min}/$dof=0.9923). The parameters $\Omega_{m0}$ and $n$ can be constrained well at $68.3\%$ confidence level by these observational data.

We also reconstructed the effective equation of state and the deceleration parameter from observational data. We found that the transition from deceleration to acceleration occurs at $z=0.95\pm0.05$. The present value of deceleration parameter was found to be $q_0=-0.3750$, meaning that the cosmic expansion has passed a maximum value (about at $z\sim 0.1$) and is now slowing down again. This is a theoretically interesting result because eternally accelerating universe (like $\Lambda$CDM) is endowed with a cosmological event horizon which prevents the construction of a conventional S-matrix describing particle interactions. Such a difficulty has been
pointed out as a severe theoretical problem for any eternally accelerated universe \citep{Hellerman:2001yi,Cline:2001nq,Carvalho:2006fy}. Some researches also indicated that a transient phase of accelerated expansion is not excluded by current observations \citep{Guimaraes:2010mw,Vargas:2011sz,Bassett:2002qu}. We note, however, it is possible to have an eternal accelerated phase and an effective equation of state crossing through $-1$ at $68.3\%$ confidence level, according to the reconstruction of the effective equation of state and the deceleration parameter. We look forward to a more comprehensive investigation including the observations of structure growth which is widely used to study $f(T)$ gravity \citep{izumi2013cosmological,chen2011cosmological,geng2013density,zheng2011growth,li2011large}, to reduce errors of the effective equation of state and the deceleration parameter at $z\sim 0$.

\begin{acknowledgements}
We thank Puxun Wu for his helpful suggestions. This research is supported by the National Natural Science Foundation of China (Grant Nos. 11235003, 11175019, 11178007, 11147028 and 11273010) and Hebei Provincial Natural Science Foundation of China (Grant No. A2011201147 and A2014201068).
\end{acknowledgements}

\bibliographystyle{raa}
\bibliography{ftmodel}

\end{document}